\documentclass[%
twocolumn,
amsmath,amssymb,
]{revtex4-2}
\usepackage{amsmath}
\usepackage{amssymb}
\usepackage{xcolor}
\usepackage[normalem]{ulem}
\usepackage{cancel}
\usepackage{graphicx}
\usepackage{enumerate}
\usepackage{enumitem}
\usepackage{bm}
\usepackage{multibib} 

\usepackage{tikz}
\usetikzlibrary{arrows.meta,positioning,decorations.pathreplacing,calc}
\usepackage{soul}
\definecolor{xcolor}{HTML}{8A0000}
\definecolor{ycolor}{HTML}{00008A}
\usepackage[colorlinks=true,allcolors=blue]{hyperref}
\begin{document}

\title{\textbf{Two-tooth bosonic quantum comb for temporal-correlation sensing} }%
\author{Shaojiang Zhu}
\email{szhu26@fnal.gov}
\affiliation{Superconducting Quantum Materials and Systems Center, Fermi National Accelerator Laboratory, Batavia, IL 60510, USA}%
\author{Xinyuan You}
\affiliation{Superconducting Quantum Materials and Systems Center, Fermi National Accelerator Laboratory, Batavia, IL 60510, USA}%
\author{Alexander Romanenko}
\affiliation{Superconducting Quantum Materials and Systems Center, Fermi National Accelerator Laboratory, Batavia, IL 60510, USA}%
\author{Anna Grassellino}
\affiliation{Superconducting Quantum Materials and Systems Center, Fermi National Accelerator Laboratory, Batavia, IL 60510, USA}%

\begin{abstract}
We characterize the causal structure of coherence transport in open bosonic systems using a two-tooth quantum comb. We show that, by mapping sequential interactions between a structured bosonic absorber and a long-lived coherent probe onto a process-tensor description, the probe functions as a temporal interferometer for environmental two-time correlations. We derive closed-form expressions for a nonmonotonic memory response arising from a competition between instantaneous thermal responsivity and the delay-dependent evolution of absorber correlations. By tuning the intertooth delay, the comb can enhance or suppress the net temperature imprint, providing direct access to population correlators and enabling discrimination of spectrally structured non-Markovian fluctuations from effective thermal noise. The protocol provides an explicit bosonic implementation of two-time thermal-correlation sensing, complementary to prior studies of memory effects in noisy quantum metrology.
\end{abstract}

\maketitle

Thermal fluctuations in quantum devices are not merely a nuisance: they are fundamental carriers of energy and information in open quantum systems~\cite{rivas2014quantum, streltsov2017colloquium, landi2024current}, and set practical limits for quantum thermometry and calorimetry, high-precision noise spectroscopy, and ultimately gate fidelity and fault tolerance in quantum processors~\cite{clerk2010introduction, kurdzialek2025quantum, barzanjeh2018manipulating, campbell2026roadmap, bylander2011noise, wang2020spectral, stace2010quantum, mallet2011quantum, krantz2019quantum}.
At the single-quantum level, ``temperature'' is an emergent dynamical imprint of discrete energy-exchange events and, crucially, their temporal correlations.

Most quantum thermometry protocols, however, rely on a single-time (``one-tooth'') interaction: a probe couples once to a thermal mode and a phase shift or dephasing envelope is read out as a snapshot of the bath occupation~\cite{de2019quantum, sharafiev2025leveraging, lvov2025thermometry, gumucs2023calorimetry, saira2016dispersive, zhu2025coherence}.
Such snapshots primarily access equal-time moments, e.g., mean occupation and instantaneous fluctuation strength, and are largely insensitive to the temporal correlators that quantify environmental memory, limiting their ability to diagnose slow, spectrally structured fluctuations.

To resolve this hidden dynamical structure, a sensing protocol must not only register noise but also record how the environment re-correlates with itself on controllable time scales~\cite{viola1999dynamical}.
In this Letter, we show that sequential interactions between a thermal absorber and a long-lived coherent probe naturally implement a two-tooth bosonic quantum comb (TBQC): a minimal, nontrivial two-time causal architecture that tracks how environmental information is acquired, stored, and re-used~\cite{chiribella2008memory, wang2023beating, chin2012quantum}. 
Building on the general metrological insight that memory or non-semigroup dynamics can improve precision scaling~\cite{smirne2016ultimate}, the TBQC scheme developed here shifts the role of memory in the sensing problem.
Rather than treating memory as a property of a noisy channel that affects estimation of an externally encoded parameter, we consider a bosonic thermal absorber whose temporal memory is itself the thermodynamic signal. 
The two-time population correlator \(\mathcal{K}(\Delta)\)
\footnote{More generally, \(\mathcal{K}=\mathcal{K}(\Delta,T)\). For compactness, we suppress the explicit \(T\)-dependence of \(\mathcal{K}\) and derived quantities when no ambiguity arises.}
is coherently transduced into the visibility of a bosonic probe through a minimal two-tooth cross-Kerr comb. 
Varying the inter-tooth delay therefore realizes a temporal interferometer for thermal population correlations, separating instantaneous thermometric responsivity from dynamical correlation responsivity.

Within a process-tensor description of the resulting effective phase-noise dynamics, we derive a compact two-time memory kernel and obtain analytic expressions for the probe visibility and the corresponding quantum Fisher information (QFI)~\cite{zhong2013fisher, liu2020quantum}. 
A central result is a \emph{non-monotonic memory response} versus the inter-tooth delay $\Delta$. 
For short delays $\Delta \ll\tau_c$, where $\tau_c$ is the absorber correlation time, preserved correlations allow the two interaction windows to interfere constructively, yielding memory-assisted thermometric precision that outperforms a single-interaction protocol under identical probe resources.
For long delays $\Delta \gg\tau_c$, the correlations vanish and the response asymptotically returns to the Markovian limit of two effectively independent interactions.
For intermediate delays $\Delta \sim\tau_c$, partially decayed correlations add dephasing without a commensurate temperature responsivity, and the QFI can dip below its memoryless (Markovian) baseline. 

This delay dependence reveals a central competition between two distinct forms of responsivity: instantaneous sensitivity to the absorber's mean thermal population and delay-dependent sensitivity to its dynamical correlations. 
By sweeping $\Delta$, the probe directly samples the thermal correlation via the memory kernel
\begin{equation}\label{eq:K}
    \mathcal{K}(\Delta)=\langle \delta n_a(t)\,\delta n_a(t+\Delta)\rangle_T,
\end{equation}
enabling a bosonic noise-spectroscopy protocol that distinguishes Markovian thermal noise from slow or spectrally structured fluctuations~\cite{das2025quantum}. 
More broadly, two-time bosonic quantum combs provide a simple and experimentally accessible route to diagnosing the causal structure of fluctuating quantum environments in circuit-QED and related solid-state platforms~\cite{altherr2021quantum, van2023device, smirne2016ultimate}.

\begin{figure}[b!]
\centering
\includegraphics[width=0.9\columnwidth, trim={4.55cm 17.2cm 4.5cm 2.03cm}, clip]{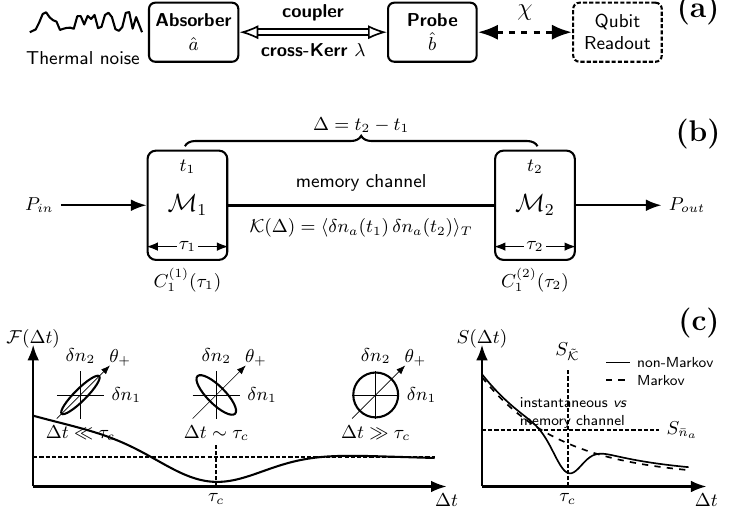}
\caption{Schematic of TBQC for thermometry and noise discrimination.
(a) A thermal absorber couples to a long-lived coherent probe via an engineered cross-Kerr interaction~$\lambda$, while the probe is monitored by a readout qubit with dispersive shift~$\chi$.
(b) Process-tensor (quantum comb) representation: the probe interacts with the absorber in two short windows $\mathcal{M}_{1,2}$ of durations $\tau_{1,2}$ centered at times $t_{1,2}$; the inter-tooth delay $\Delta$ exposes the absorber’s population correlator $\mathcal{K}(\Delta)$ that mediates memory between the interactions.
(c) Schematic delay dependence of the temperature information $\mathcal{F}(\Delta)$, with insets illustrating the evolution of joint number-fluctuation correlations from strongly elongated (short delay) to isotropic (long delay). Right: the competing contributions of an instantaneous responsivity $S_{\bar n_a}$ and a decaying memory channel $S_{\tilde{\mathcal{K}}}$ set the observed non-monotonic response.}
\label{fig:TBQC_schematic}
\end{figure}

We consider a TBQC scheme built on the coherence-mediated thermometry architecture~\cite{zhu2025coherence}, as sketched in Fig.~\ref{fig:TBQC_schematic}(a). In this architecture, a thermalized absorber mode $\hat{a}$ couples dispersively to a long-lived coherent probe mode $\hat{b}$ through the engineered cross-Kerr interaction \(H_{\rm int}=\lambda \hat{a}^\dagger \hat{a}\hat{b}^\dagger \hat{b}\), such that shot-to-shot absorber number fluctuations imprint a temperature-dependent phase on the probe while largely preserving probe energy.
The probe is then monitored through a dispersively coupled readout qubit, which converts the probe phase-space displacement into an experimentally accessible signal without requiring the qubit to remain strongly coupled to the noisy absorber throughout the sensing interval. As such, the two-tooth protocol requires no additional hardware element relative to Ref.~\cite{zhu2025coherence}; the key added requirement is preserving and reading out the probe coherence over a controlled inter-tooth delay.

Fig.~\ref{fig:TBQC_schematic}(b) and (c) recast this sequence as a TBQC: the probe interacts with the absorber in two short windows $\mathcal{M}_{1,2}$ with durations $\tau_{1,2}$ separated by a controllable delay $\Delta=t_2-t_1$. 
Operationally, each window is one probe ``tooth'', namely a calibrated interval during which the effective cross-Kerr coupling is active. The protocol prepares the probe in a coherent state, applies the first tooth, lets the absorber evolve freely for time \(\Delta\) while the probe retains the imprinted phase, applies the second tooth, and finally reads out the probe coherence.
Varying $\Delta$ coherently compares the phase kicks accumulated in the two windows, mapping connected population correlations onto phase and visibility signatures inaccessible to any single-snapshot scheme~\cite{gao2018programmable}.
As the delay crosses the absorber correlation time $\tau_c$, the joint fluctuation statistics evolve from strongly correlated to effectively independent (elliptical insets), yielding a characteristic, non-monotonic dependence of the QFI that can be interpreted as a competition between an instantaneous responsivity channel and a decaying memory channel.

\begin{figure*}[t!]
\centering
\includegraphics[width=1.95\columnwidth, trim={0.1cm 0.3cm 0.1cm 0.1cm}, clip]{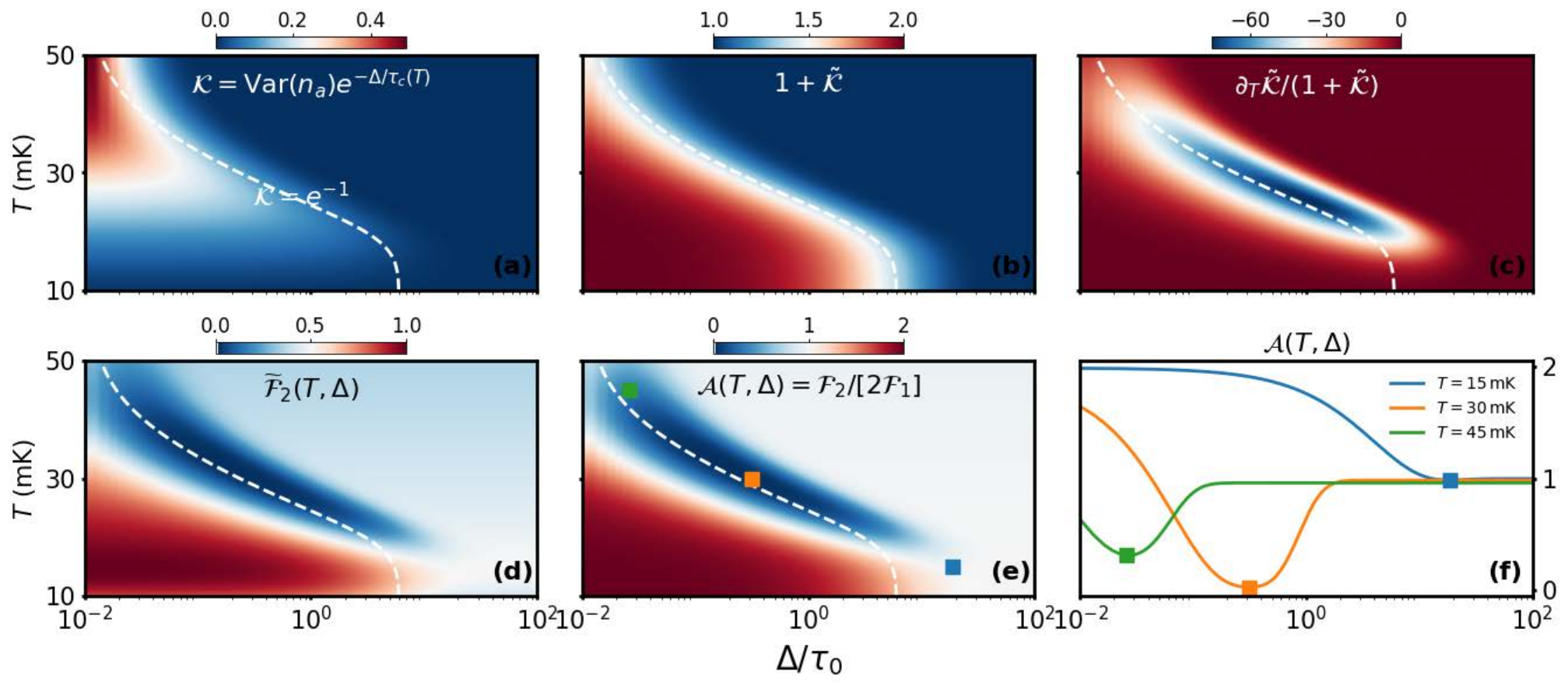}
\caption{Memory kernel and non-monotonic memory response of the two-tooth comb.
(a) Lorentzian correlation kernel (Eq.~\eqref{eq:Lor_K}) showing the exponential correlation envelope and its temperature-dependent decay.  
The white dashed curve denotes $\tilde{\mathcal{K}}=e^{-1}$ (\textit{i.e.}, $\Delta=\tau_c(T)$).  
(b) Correlation-induced variance factor $1+\tilde{\mathcal{K}}$, which quantifies how temporal correlations increase the probe phase variance.   
(c) Relative responsivity $\partial_T \tilde{\mathcal{K}}/(1+\tilde{\mathcal{K}})$, revealing the locus of reduced temperature sensitivity and its competition with the population responsivity. 
(d) Normalized two-tooth QFI \(\widetilde{\mathcal{F}}_2=\mathcal{F}_2/\max(\mathcal{F}_2)\), emphasizing the non-monotonic structure of the kernel responsivity and correlation decay.
(e) Corresponding memory efficiency $\mathcal{A}=\mathcal{F}_2/[2\mathcal{F}_1]$, showing enhancement ($\mathcal{A}>1$) at short delays where correlations dominate and suppression ($\mathcal{A}<1$) where temperature responsivities compete.  
Colored markers indicate the delay at which $\mathcal{A}$ reaches its minimum for selected temperatures. 
(f) Line cuts of $\mathcal{A}(\Delta)$ at three representative temperatures (15, 30, and 45 mK).  
At low temperature, the minimum of $\mathcal{A}$ is shifted away from the $1/e$ contour due to the weak temperature dependence of the correlation time, whereas at higher temperatures the minimum aligns with the correlation-time boundary.}
\label{fig:process_kernel}
\end{figure*}

\paragraph*{Memory kernel:}
Within the Gaussian cross-Kerr model, the resulting two-tooth coherence takes the compact form in the weak-dephasing regime~\cite{szhu_suppl}:
\begin{equation}\label{eq:C2}
    C_2(\Delta) = C_1^{(1)}(\tau_1)\,C_1^{(2)}(\tau_2)\,\exp\!\big[-2\lambda^2\tau_1\tau_2\,\mathcal{K}(\Delta)\big],
\end{equation}
where the single-interaction coherence for tooth \(j\) is
\begin{equation}\label{eq:C1}
    C_1^{(j)}(\tau_j)=e^{-(\lambda \tau_j)^2\bar{n}_a}, \quad j\in\{1,2\}.
\end{equation}
The absorber follows a fluctuating population trajectory \(n_a(t) = \bar{n}_a + \delta n_a(t)\), and the two interaction windows sample this trajectory at times \(t_1\) and \(t_2=t_1+\Delta\).

Eq.~\eqref{eq:C2} makes the memory physics explicit.
The product $C_1^{(1)}\,C_1^{(2)}$ is the benchmark obtained by probing two independent noise realizations, while the exponential kernel factor encodes how strongly the two interactions are correlated in time.  
For slow absorber dynamics, $\mathcal{K}(\Delta)\!\approx\!\mathcal{K}(0)$ when $\Delta \ll \tau_c$, both teeth interrogate nearly the same fluctuation history; the correlated term then contributes additional dephasing and reduces $C_2(\Delta)$ relative to the independent benchmark.
At the same time, this correlation channel carries temperature dependence through $\partial_T\mathcal{K}$, so the temperature slope, and hence the QFI, can be enhanced even when the visibility is reduced.
For intermediate delays $\Delta\sim\tau_c$, the kernel has decayed but remains finite, so the memory term can add contrast loss without a comparable gain in temperature responsivity; in this regime the memory efficiency can fall below the independent benchmark.
In the Markovian limit $\Delta \gg \tau_c$, $\mathcal{K}\!\to\!0$ and $C_2\!\to\!C_1^{(1)}C_1^{(2)}$, showing that the two-tooth comb smoothly reduces to a memoryless Ramsey-like sequence~\cite{smirne2016ultimate}. 

The behavior of the memory kernel underlying the two-tooth protocol is summarized in Fig.~\ref{fig:process_kernel}(a)--(c). Here, we assume an exponentially decaying kernel,
\begin{equation}\label{eq:Lor_K}
    \mathcal{K}(\Delta) = \mathrm{Var}(n_a)\,\tilde{\mathcal{K}}(\Delta), \quad \tilde{\mathcal{K}}=e^{-\Delta / \tau_c(T)},
\end{equation}
which corresponds to a Lorentzian absorber noise spectrum with correlation time $\tau_c(T)$~\cite{mickelsen2023effects, de2014fluctuations}.
Panel~(a) shows how the kernel evolves with $T$ and $\Delta$, highlighting the exponential decay of thermal photon-number correlations at each temperature. 
The white dashed curve marks the contour $\tilde{\mathcal{K}}=e^{-1}$, i.e., $\Delta=\tau_c$, where correlations have dropped to \(1/e\) of their zero-delay value.
Panel~(b) displays the amplitude factor $(1+\tilde{\mathcal{K}})$ appearing in Eq.~\eqref{eq:decomposed_A}, which quantifies the correlation-induced increase of the phase variance.
This correlated contribution is broad at low temperature but narrows as the absorber correlation time decreases with increasing $T$.
Panel~(c) shows the relative kernel responsivity $\partial_T (1 + \tilde{\mathcal{K}})/(1+\tilde{\mathcal{K}})$, which controls how memory modifies the temperature slope entering the QFI.
Near the \(1/e\) contour, \(\partial_T(1+\tilde{\mathcal{K}})\) becomes largest in magnitude and develops a pronounced negative ridge, reflecting competition with the population responsivity.
We will show that this competition can locally suppress the memory efficiency below the independent-tooth benchmark.

\paragraph*{Memory efficiency:}
Using the coherence envelopes in Eqs.~\eqref{eq:C1} and~\eqref{eq:C2}, we compute the QFI associated with the specified one- and two-tooth probe sequences~\cite{szhu_suppl}. 
For the two-tooth protocol, the temperature response of the dephasing exponent separates into
\begin{equation}\label{eq:dGamma_phi2}
\partial_T \Gamma_{\phi_2}=\lambda^2\big[(\tau_1^2+\tau_2^2)\,\partial_T \bar n_a+2\tau_1\tau_2\,\partial_T \mathcal{K}(\Delta)\big],
\end{equation}
where the first term is the additive response expected from two independent teeth, while the second term is the delay-dependent response of the memory kernel. Thus, temporal correlations open an additional thermometric channel, carried by \(\partial_T\mathcal{K}(\Delta)\), beyond instantaneous number fluctuations alone.

To quantify this contribution, we define the memory efficiency
\begin{equation}\label{eq:A}
    \mathcal{A}(\Delta)=\frac{\mathcal{F}_2(\Delta)}{\mathcal{F}_1^{(1)}(\tau_1)+\mathcal{F}_1^{(2)}(\tau_2)} .
\end{equation}
Here \(\mathcal{F}_1^{(j)}\) and \(\mathcal{F}_2\) are protocol-level QFIs for the specified one- and two-tooth output probe states, optimized only over measurements on the final probe state~\cite{braunstein1994statistical,paris2009quantum}. They should not be interpreted as ultimate bounds optimized over arbitrary input states, adaptive controls, comb protocols, or joint probe--absorber measurements~\cite{escher2011general, chiribella2009theoretical}.
In the weak-dephasing limit, this definition is normalized so that \(\mathcal{A}\to1\) in the absence of temporal correlations, while deviations from unity quantify the gain or penalty induced by memory.

For a practical implementation of \(N\) repeated measurements, the Cram\'er--Rao bound gives the minimum achievable temperature uncertainty, \(\delta T\ge 1/\sqrt{N\mathcal{F}}\). 
Thus, the memory efficiency \(\mathcal{A}\) provides a direct metrological measure of the temporal-correlation gain: it is the protocol-level QFI ratio relative to independent-tooth sampling, so \(\sqrt{\mathcal{A}}\) gives the corresponding precision scaling for the same number of repetitions.
From Eq.~\eqref{eq:A}, \(\mathcal{A}>1\) means that the two-tooth protocol requires fewer repetitions than the independent-tooth benchmark to reach the same temperature uncertainty, whereas \(\mathcal{A}<1\) indicates a correlation-induced penalty.

Using the same kernel model in Eq.~\eqref{eq:Lor_K} and taking two identical teeth ($\tau_1 = \tau_2$), we evaluate the two-tooth $\mathcal{F}_2$ and the corresponding $\mathcal{A} = \mathcal{F}_2 / 2\mathcal{F}_1$. The results are summarized in Fig.~\ref{fig:process_kernel}(d)--(f).
Notably, both $\mathcal{F}_2$ and $\mathcal{A}$ exhibit a characteristic non-monotonic behavior, showing either enhancement or suppression depending on the absorber temperature and the separation between the two teeth [see Panels~(d) and (f)].
This behavior arises from a competition between instantaneous population responsivity and delay-dependent correlation responsivity.

To make this competition explicit, we rewrite Eq.~\eqref{eq:A} in the weak-dephasing limit~\cite{szhu_suppl}
\begin{equation}\label{eq:decomposed_A}
    \mathcal{A} = \underbrace{(1 + \tilde{\mathcal{K}})}_{\small \text{Variance Gain}} \times \underbrace{\left[ 1 + \frac{S_{\tilde{\mathcal{K}}}}{S_{\bar{n}_a}} \right]^2}_{\small \text{Responsivity Competition}},
\end{equation}
in which the prefactor $(1 + \tilde{\mathcal{K}}) \geq 1$ acts as a gain knob that quantifies how correlated fluctuations increase the total phase variance, while the bracketed factor encodes the competition between two temperature responsivities: the population responsivity
$S_{\bar{n}_a} = \partial_T \bar{n}_a / \bar{n}_a$ and the correlation responsivity $S_{\tilde{\mathcal{K}}} = \partial_T(1 + \tilde{\mathcal{K}}) / (1+\tilde{\mathcal{K}})$.
When $S_{\tilde{\mathcal{K}}}$ reinforces $S_{\bar{n}_a}$, memory sharpens the temperature slope and enhances $\mathcal{F}_2$ ($\mathcal{A} > 1$).
When $S_{\tilde{\mathcal{K}}}$ partially cancels $S_{\bar{n}_a}$, the net slope is reduced and correlations can degrade $\mathcal{F}_2$, driving $\mathcal{A}$ below unity ($\mathcal{A} < 1$)~\cite{chin2012quantum}.
Panel~(f) illustrates the temperature dependence of this competition through the corresponding variation of the noise correlation time $\tau_c$.

We emphasize that the mechanism is not tied to the particular exponential kernel in Eq.~\eqref{eq:Lor_K}: for any correlated noise source whose dephasing can be decomposed into a diagonal (population) term and a two-time kernel, the relative size and sign of the two responsivities $S_{\bar{n}_a}$ and $S_{\tilde{\mathcal{K}}}$ in Eq.~\eqref{eq:decomposed_A} determine whether correlations enhance or suppress the protocol-level thermometric performance~\cite{smirne2016ultimate}.

The two-tooth sequence is not claimed to be globally optimal, but rather is a minimal, nontrivial element of a systematic multi-tooth comb hierarchy. 
It is the simplest architecture that accesses a genuine temporal correlation of the absorber and is sufficient for the central objective of separating instantaneous thermometric responsivity from delay-dependent correlation responsivity. 

For a general \(m\)-tooth protocol with equally spaced, identical teeth, the temperature response of the dephasing rate becomes~\cite{szhu_suppl}
\begin{equation*}
\partial_T\Gamma_{\phi_m}\simeq\lambda^2\tau^2\Big[m\,\partial_T\bar n_a
+2\sum_{r=1}^{m-1}(m-r)\partial_T\mathcal{K}(r\Delta)\Big].
\end{equation*}
Thus, the \(m\)-tooth comb does not simply repeat the two-tooth measurement. 
It reshapes the relative weight of the equal-time response and the temporal-memory response through the pair-counting factors \(m-r\). 
Slowly decaying memory can be enhanced by the many correlated tooth pairs, whereas rapidly decaying or sign-changing responsivity can be averaged out or partially canceled. 
This additional control, however, comes at the cost of longer interrogation times and more demanding calibration.

\begin{figure}[t!]
\centering
\includegraphics[width=1.0\columnwidth]{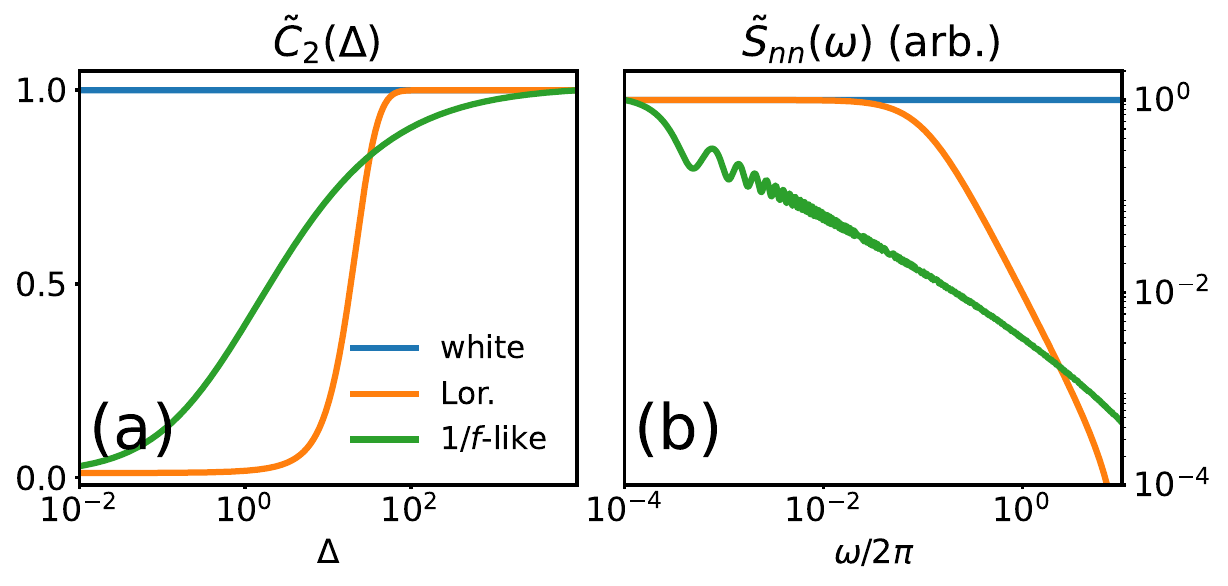}
\caption{(a) Normalized two-tooth coherence $\tilde{C}_2(\Delta)$ for white, Lorentzian, and $1/f$-like population noise. (b) Corresponding normalized spectra $\tilde{S}_{nn}(\omega)$ reconstructed from $\tilde{C}_2(\Delta)$.}
\label{fig:noise_spectrum_discrimination}
\end{figure}

\paragraph*{Correlated Noise Spectra.}
The memory kernel $\mathcal{K}$ in Eq.~\eqref{eq:K} can be identified with the absorber population autocorrelation: by sweeping the inter-tooth delay $\Delta$, one modulates the two-tooth probe coherence $C_2(\Delta)$ in Eq.~\eqref{eq:C2} and thereby directly samples $\mathcal{K}(\Delta)$~\cite{alvarez2011measuring, paz2014general}. 

According to the Wiener--Khinchin relation~\cite{brockwell2002introduction},
\begin{equation}\label{eq:WienerKhinchin}
    \mathcal{K}(\Delta) = \int_{-\infty}^{\infty}\!\frac{d\omega}{2\pi}\,S_{nn}(\omega)\,e^{-i\omega\Delta},
\end{equation}
the measured delay dependence of $C_2(\Delta)$ provides access to the absorber number-fluctuation spectrum \(S_{nn}(\omega)\), subject to the usual finite-window and calibration limitations.

In filter-function language, the two-tooth sequence acts as a comb-like spectral weight that includes an oscillatory factor $\sim \cos(\omega\Delta)$, so varying $\Delta$ shifts the interference fringes across frequency and reshapes which components of $S_{nn}(\omega)$ contribute to dephasing~\cite{clerk2010introduction, viola1999dynamical, paz2014general}. 
Under calibrated finite-window conditions, delay-domain visibility data can distinguish flat, finite-bandwidth, and low-frequency-dominated population noise. 
This suggests a practical \emph{in situ} diagnostic use of the same bosonic mode for correlated fluctuations in cryogenic detectors and quantum sensors, including slow fluctuators, quasiparticle bursts, and radiation-induced noise~\cite{wang2020spectral,mazin2009microwave,vepsalainen2020impact,sefkow2016experimental}.

Figure~\ref{fig:noise_spectrum_discrimination} illustrates this discrimination. 
Panel~(a) plots the normalized coherence $\tilde{C}_2(\Delta)$: white noise gives a nearly flat response, Lorentzian noise shows a crossover on the correlation-time scale, and $1/f$-like noise yields a slow rolloff. 
Panel~(b) shows the corresponding spectra $\tilde{S}_{nn}(\omega)$, highlighting flat (white), finite-bandwidth (Lorentzian), and low-frequency--dominated ($1/f$-like) behavior.

\paragraph*{Summary}
We introduced a TBQC protocol that repurposes memory in bosonic thermometry: rather than treating temporal correlations as a feature of a noisy channel, the protocol makes the absorber's memory itself the thermodynamic signal. 
Two separated cross-Kerr interaction windows coherently transduce the absorber's two-time population correlator \(\mathcal{K}(\Delta)\) into the measured probe visibility. 
This mapping separates equal-time thermometric responsivity from delay-dependent correlation responsivity and yields a finite-delay competition in which memory can either enhance or suppress the protocol-level QFI.

The two-tooth sequence is the minimal nontrivial member of a systematic multi-tooth comb hierarchy. 
Higher-tooth extensions sample weighted combinations of absorber correlations and therefore act as temporal filters for bosonic population noise, with potential gains in correlation selectivity and with practical tradeoffs in sequence duration, probe coherence, timing stability, and calibration~\cite{szhu_suppl}.

In circuit-QED, the TBQC scheme extends the coherence-mediated thermometry architecture~\cite{zhu2025coherence} by replacing one interaction window with two separated cross-Kerr windows; the main added requirement is preserving and reading out probe coherence over the inter-tooth delay, rather than introducing new hardware. 
Importantly, these added requirements enter primarily as calibrated contrast loss, timing uncertainty, or averaging overheads; within the calibrated operating window, they reduce the attainable sensitivity but do not compromise the correlation-filtering function of the comb~\cite{szhu_suppl}. 
This provides a correlation-resolved route to bosonic thermometry and population-noise spectroscopy in structured thermal environments, with potential applications to diagnosing spurious heating, energy-flow pathways, and engineered reservoirs in circuit-QED and other hybrid bosonic platforms.

\paragraph*{Acknowledgments.}
This work was supported by the U.S. Department of Energy, Office of Science, National Quantum Information Science Research Centers, Superconducting Quantum Materials and Systems Center (SQMS), under Contract No. 89243024CSC000002. Fermilab is managed by FermiForward Discovery Group, LCC, acting under Contract No. 89243024CSC000002.

\paragraph*{Data availability—} The data that support the findings of this article are openly available~\footnote{Data and python script available at: \url{https://github.com/sjzhuster/cmqt-code.git}, embargo periods may apply.}

\appendix
\section{Coherence Envelope}\label{app:coherence_envelope}
When the accumulated phase noise is well described by Gaussian diffusion, the dephased probe state can be written as the phase-averaged mixture
\begin{equation*}
    \rho_b(T) = \int d\phi\,P(\phi)\,|\alpha e^{i\phi}\rangle\!\langle\alpha e^{i\phi}|,
\end{equation*}
with Gaussian distribution \(P(\phi)=\tfrac{e^{-\phi^2/2\sigma_\phi^2}}{\sqrt{2\pi\sigma_\phi^2}}.\)
The coherence envelope is the overlap of the initial coherent state with the diffused mixture:
\begin{equation*}
    C(\tau) = \langle \alpha | \rho_b(T) | \alpha \rangle = \int d\phi\, P(\phi)\,\big|\langle \alpha | \alpha e^{i\phi}\rangle\big|^2,
\end{equation*}
and can be evaluated exactly:
\begin{equation}\label{eq:SI_C}
    C(\tau) = \exp\Big[-2|\alpha|^2 \big(1-e^{-\sigma_\phi^2/2}\big)\Big]. 
\end{equation}
with \(\sigma_\phi^2 = \mathrm{Var}(\phi_b)\) being the coherent phase variance. For simplicity, we set the photon number in the probe $|\alpha|^2 = 1$ in the following discussions.

\paragraph*{One-tooth coherence envelope.}
We consider a coherent probe mode initially prepared in $|\alpha\rangle$, coupled dispersively to a thermal absorber mode $a$ through the cross-Kerr interaction \(H_{\mathrm{int}} = \lambda\, \hat n_a \hat n_b\), where $\lambda$ is the cross-Kerr rate.  
During an interaction of duration $\tau$, the probe accumulates a phase conditioned on the instantaneous thermal photon number,
\begin{equation*}
    |\alpha\rangle \rightarrow |\alpha e^{i\phi_b}\rangle, \quad \phi_b = \lambda\, n_a \tau.
\end{equation*}
Fluctuations of $n_a$ in a thermal state therefore induce random probe phases.

For a thermal mode with mean occupation \(\bar n_a(T) = \big[\exp(\hbar \omega_a / k_B T) - 1\big]^{-1},\) the number fluctuations is \(\mathrm{Var}(n_a) = \bar n_a(1+\bar n_a).\) In the limit \(\bar{n}_a \ll 1\) at low temperatures, \(\mathrm{Var}(n_a) \approx \bar n_a\) and we keep this approximation in the work.

The induced phase therefore has variance
\begin{equation}\label{eq:SI_phase1}
    \sigma_\phi^2 = \mathrm{Var}(\phi_b) = (\lambda\tau)^2\, \bar n_a.
\end{equation}
Substituting Eq.~\eqref{eq:SI_phase1} in Eq.~\eqref{eq:SI_C} gives the one-tooth coherence envelope:
\begin{equation}\label{eq:SI_C1}
    C_1(\tau) = \exp\!\big[-2(1-e^{-\tfrac{1}{2}\Gamma_{\phi_1}})\big],
\end{equation}
with the effective dephasing rate \(\Gamma_{\phi_1} = (\lambda\tau)^2\, \bar n_a\).
In the weak dephasing regime \(\Gamma_{\phi_1} \ll 1\), Eq.~\eqref{eq:SI_C1} is simplified as
\begin{equation}\label{eq:SI_C1_WD}
    C_1(\tau) = \exp\!\big[-(\lambda\tau)^2\, \bar n_a\big],
\end{equation}
and its first-order approximation is \(C_1 \approx 1 - (\lambda\tau)^2\, \bar n_a.\)

\paragraph*{Two-tooth coherence envelope.}
We now extend the one-tooth analysis to the case of two-tooth interactions.  
The coherent probe $b$ interacts dispersively with the thermal absorber $a$ during two windows of durations $\tau_1$ and $\tau_2$, centered at times $t_1$ and $t_2=t_1+\Delta$, respectively.  
During each window, the dynamics are governed by the same cross-Kerr Hamiltonian: \(H_{\mathrm{int}} = \lambda\, \hat n_a \hat n_b.\)

For a fixed trajectory of the absorber occupation, the probe acquires a phase
\begin{equation*}
    \phi_{2} = \lambda\big[\tau_1 n_a(t_1) + \tau_2 n_a(t_2)\big],
\end{equation*}
so that an initial coherent state $|\alpha\rangle$ is mapped to $|\alpha e^{i\phi_{2}}\rangle$ after both interactions.  
The absorber fluctuations render $\phi_2$ a random variable.
The deterministic mean phase is removed by demodulation, so the coherence envelope is governed by the phase variance.

We assume that the absorber number fluctuations are stationary and that the relevant statistics are well captured by their second moments. 
Therefore, the population autocorrelation function at delay $\Delta$ and fixed temperature $T$ can be expressed as: 
\begin{equation}\label{eq:SI_K}
    \mathcal{K}(\Delta) = \langle \delta n_a(t_1)\,\delta n_a(t_2)\rangle,
\end{equation}
with $\delta n_a(t)=n_a(t)-\bar n_a$. 
The two-tooth phase variance Var($\phi_2$) reads
\begin{align}\label{eq:SI_phase2}
    \sigma_{\phi_2}^2 = \lambda^2\Big[(\tau_1^2 + \tau_2^2)\,\bar n_a + 2 \tau_1\tau_2\,\mathcal{K}(\Delta) \Big].
\end{align}
Here, $\mathcal{K}(\Delta)$ encodes the temporal correlations of the population fluctuations.

Substituting Eq.~\eqref{eq:SI_phase2} into Eq.~\eqref{eq:SI_C} yields the two-tooth coherence envelope
\begin{equation}\label{eq:SI_C2}
    \scalebox{0.9}{$C_2(\Delta) = \exp\Big\{-2\Big[1-e^{-\tfrac{1}{2}\lambda^2\big[(\tau_1^2 + \tau_2^2)\bar{n}_a + 2 \tau_1\tau_2\,\mathcal{K}(\Delta) \big]}\Big]\Big\}.$}
\end{equation}
In the weak-dephasing regime, Eq.~\eqref{eq:SI_C2} reduces to Eq.~\eqref{eq:C2} in the main text:
\begin{equation}\label{eq:SI_C2_WD}
\begin{aligned}
    C_2(\Delta)
    &=\exp\!\left\{-\lambda^2\left[(\tau_1^2+\tau_2^2)\bar n_a + 2\tau_1\tau_2\,\mathcal{K}(\Delta)
    \right]\right\} \\
    &=C_1^{(1)}(\tau_1)\,C_1^{(2)}(\tau_2)\,\exp\!\left[-2\lambda^2\tau_1\tau_2\,\mathcal{K}(\Delta)\right],
\end{aligned}
\end{equation}
with the effective two-tooth dephasing rate \[\Gamma_{\phi_2} = \lambda^2\big[(\tau_1^2 + \tau_2^2)\bar{n}_a + 2 \tau_1\tau_2\,\mathcal{K}(\Delta)\big].\]
For slow absorber dynamics, \(\mathcal{K}(\Delta)\approx \bar n_a\) in the low-occupation limit, so the two teeth sample nearly the same fluctuation history and acquire maximal correlated dephasing. 
For fast absorber dynamics, \(\mathcal{K}(\Delta)\to0\), and \(C_2\) reduces to the product of two independent one-tooth coherences.
The first-order expansion is
\[C_2(\Delta)\approx1-\lambda^2\left[(\tau_1^2+\tau_2^2)\bar n_a+2\tau_1\tau_2\,\mathcal{K}(\Delta)
\right].\]

\section{Quantum Fisher Information (QFI)}

We consider a qubit probe initialized in the superposition \(|+\rangle=(|0\rangle+|1\rangle)/\sqrt{2}\).  
Temperature-dependent fluctuations in the coupled environment induce pure dephasing during an interaction of duration $\tau$, producing the reduced density matrix
\begin{equation*}\rho = \frac{1}{2}
\begin{pmatrix}
1 & C \\
C & 1
\end{pmatrix},\end{equation*}
where $C$ is the temperature-dependent coherence envelope (Ramsey visibility). 
Any deterministic phase is removed by standard demodulation, so the QFI considered here is the coherence-envelope contribution associated with temperature-dependent dephasing.

The corresponding Bloch vector is 
\begin{equation*}
\vec{\mathbf{r}}=\big(C,\,0,\,0\big),\qquad r=|\mathbf r|=C.
\end{equation*}
For any qubit state undergoing pure dephasing, the QFI for estimating $T$ reduces to a single “radial’’ term~\cite{liu2020quantum,zhong2013fisher}:
\begin{equation}\label{eq:SI_F}
\mathcal{F}_T = \frac{\big(\partial_T C\big)^2}{1-C^2}.
\end{equation}
This expression reflects the fact that all temperature information is encoded solely in the decay of coherence; there is no phase-sensitive contribution because the eigenvectors of $\rho(T)$ are temperature independent.

The same expression applies directly to the bosonic probe used in the two-tooth comb approach.  
After one or two interaction windows, the probe’s reduced state is characterized by a complex visibility whose phase is temperature independent; thus only its magnitude $C$ carries temperature information.

The QFI used here should be understood as a protocol-level benchmark. 
It is the QFI of the reduced output probe state generated by a specified one- or two-tooth sequence, after tracing over the absorber and its bath. 
Thus Eq.~\eqref{eq:SI_F} is optimized over measurements on this final probe state, but not over arbitrary probe preparations, joint probe--absorber measurements, adaptive controls, or all possible multi-tooth combs. 
The memory efficiency defined below therefore quantifies a correlation-induced improvement within the specified TBQC readout, not an unconditional quantum-metrological scaling advantage.

\paragraph*{One-tooth QFI.}
Differentiating Eq.~\eqref{eq:SI_C1_WD} and substituting into Eq.~\eqref{eq:SI_F}, we get the one-tooth coherence QFI
\begin{equation}\label{eq:SI_F1}  
    \mathcal{F}_1(\tau) = \frac{(\partial_T \Gamma_{\phi_1}\,e^{-\Gamma_\phi})^2 C_1^2}{1-C_1^2},
\end{equation}
with \(\partial_T \Gamma_{\phi_1} = (\lambda\tau)^2\bar n_a \partial_T\,\bar n_a.\)
This expression characterizes the optimal temperature sensitivity obtainable from the output probe state of a single coherence-mediated sampling of the absorber.

\paragraph*{Two-tooth QFI.}
Differentiating the two-tooth coherence envelope Eq.~\eqref{eq:SI_C2_WD} and substituting into Eq.~\eqref{eq:SI_F}, we have the full expression of the two-tooth QFI
\begin{equation}\label{eq:QFI_two_tooth_final}
    \mathcal{F}_2(\Delta) = \frac{\Big[e^{-\Gamma_{\phi_2}}\,\partial_T \Gamma_{\phi_2}\Big]^2C_2^2}{1-C_2^2},
\end{equation}
with \(\partial_T \Gamma_{\phi_2} = \lambda^2(\tau_1^2+\tau_2^2)\partial_T n_a + 2 \lambda^2\tau_1\tau_2\partial_T \mathcal{K}\).

We define the memory efficiency as
\begin{equation}\label{eq:SI_A}
    \mathcal{A}(T,\Delta) = \frac{\mathcal{F}_2}{\mathcal{F}_1^{(1)} + \mathcal{F}_1^{(2)}},
\end{equation}
or equivalently as a difference
\begin{equation*}
    \mathcal{F}_{\mathrm{mem}} = \mathcal{F}_2 - (\mathcal{F}_1^{(1)} + \mathcal{F}_1^{(2)}).
\end{equation*}
This framework captures how population correlations modify the effective temperature sensitivity of the probe, allowing either enhancement or suppression relative to independent one-tooth sampling.

\section{Correlated Noise Spectra}
The absorber’s temporal correlations are encoded in $\mathcal{K}(\Delta)$, which is related to the noise spectrum via the Wiener--Khinchin theorem:
\begin{equation*}\label{eq:SI_WK}
    \mathcal{K}(\Delta) = \frac{1}{2\pi}\int_{-\infty}^{\infty} S_{nn}(\omega)\,e^{-i\omega\Delta}\,d\omega .
\end{equation*}

Thus, calibrated measurements of $C_2(\Delta)$ provide access to $\mathcal{K}(\Delta)$ and, within the finite delay window, to the corresponding spectrum $S_{nn}(\omega)$.
Here, we assume the probe mode remains coherent over the relevant delay window, or that its calibrated envelope has been divided out, and the only source of dephasing is the absorber mode associated with memory kernel $\mathcal{K}_a(\Delta)$.
Isolating the correlated contribution gives
\begin{equation*}
    \ln C_2(\Delta) + \lambda^2(\tau_1^2+\tau_2^2)\bar n_a = -2\lambda^2\tau_1\tau_2\,\mathcal{K}(\Delta).
\end{equation*}

The autocorrelation function can therefore be reconstructed as
\begin{equation}\label{eq:SI_K_from_visibility}
    \mathcal{K}(\Delta) = -\frac{1}{2\lambda^2\tau_1\tau_2}\Big[\ln C_2(\Delta)+  \lambda^2(\tau_1^2+\tau_2^2)\bar n_a\Big].
\end{equation}

Fourier transforming Eq.~\eqref{eq:SI_K_from_visibility} yields the reconstructed spectrum:
\begin{equation}\label{eq:SI_Sn_reconstruction}
    S_{nn}(\omega) = 2\int_{0}^{\infty}\mathcal{K}(\Delta)\,\cos(\omega\Delta)\,d\Delta .
\end{equation}
In practice, the upper integration limit is set by the accessible delay range, so the reconstructed spectrum has the usual finite-window resolution.

\section{Kernel Model and Memory Efficiency}\label{app:fig2_construction}
\begin{figure*}[t!]
\centering
\includegraphics[width=0.99\linewidth]{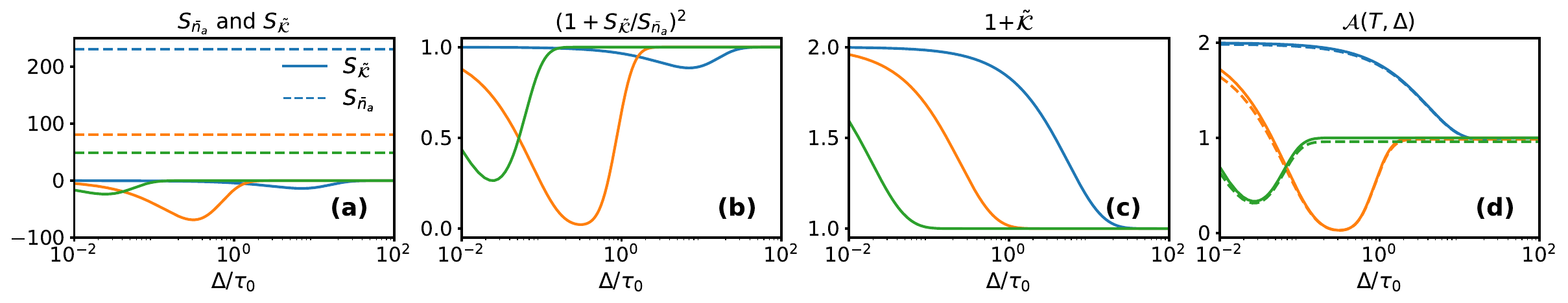}
\caption{
Competition between population and correlation responsivities in the two-tooth memory efficiency.
(a) Line cuts of the absorber population responsivity $S_{\bar{n}_a}$ (dashed) and the correlation responsivity $S_{\tilde{\mathcal{K}}}$ (solid) as a function of delay $\Delta$ for selected temperatures: 15 mK (blue), 30 mK (yellow), and 45 mK (green).
(b) Responsivity factor $(1+S_{\tilde{\mathcal{K}}}/S_{\bar{n}_a})^2$ entering the approximate expression for the memory efficiency.
(c) Amplitude gain $1+\tilde{\mathcal{K}}$, illustrating the decay of temporal correlations with increasing delay.
(d) Approximate memory efficiency $\mathcal{A} \simeq (1+\tilde{\mathcal{K}})(1+S_{\tilde{\mathcal{K}}}/S_{\bar{n}_a})^2$ for the same temperatures, showing regions of enhancement ($\mathcal{A}>1$) and suppression ($\mathcal{A}<1$) arising from the competition between thermal population responsivity and kernel responsivity. 
Dashed lines are reproduced from Fig.~\ref{fig:process_kernel}(f).}
\end{figure*}

In Fig.~\ref{fig:process_kernel}, we illustrate how a temperature-dependent absorber memory kernel shapes both the two-tooth QFI and the associated memory efficiency.
Here we summarize the kernel model, the explicit expressions used to compute the plotted quantities, and the physical content of each panel.

We consider a thermal absorber with photon-number fluctuations characterized by an exponential temporal autocorrelation, corresponding to a Lorentzian noise spectrum,
\begin{equation}\label{eq:SI_Lor_K}
    \mathcal{K}(\Delta) = \bar{n}_a\,e^{-\Delta/\tau_c(T)},  
\end{equation}
where $\Delta$ is the time separation between the two interaction windows, $\bar{n}_a(T)$ is the thermal photon number, and $\tau_c(T)$ is a temperature-dependent correlation time.
For concreteness we take a smooth crossover between a long correlation time $\tau_{\max}$ at low temperature and a short correlation time $\tau_{\min}$ at high temperature,
\begin{equation*}
    \tau_c(T) = \frac{\tau_{\max} + \tau_{\min}\,r(T)}{1 + r(T)}, \quad r(T) = \bigg(\frac{T}{T_c}\bigg)^{\gamma},
\end{equation*}
where $T_c$ sets the crossover temperature and $\gamma$ controls the sharpness of the transition.

In Fig.~\ref{fig:process_kernel} we fix \(\tau_{\max} = 6\,\tau_0,~\tau_{\min} = 0.01\,\tau_0,~T_c = 20~\mathrm{mK},~\mathrm{and}~\gamma = 8\), and scan temperatures $10~\text{mK} \leq T \leq 50~\text{mK}$ and delays $10^{-2}\, \tau_0 \leq \Delta \leq 10^{2}\,\tau_0$ on a logarithmic grid, with $\tau_0$ being a reference time used only to render the delay axis dimensionless.

To isolate the temporal structure of the correlations, we work with the normalized kernel
\begin{equation*}
    \tilde{\mathcal{K}}(\Delta) = \frac{\mathcal{K}(\Delta)}{\mathcal{K}(0)} = e^{-\Delta/\tau_c(T)},
\end{equation*}
and characterize the temperature sensitivity of the memory kernel itself with the defined kernel responsivity
\begin{equation*}
S_{\tilde{\mathcal{K}}}(\Delta)=\frac{\partial_T \tilde{\mathcal{K}}(\Delta)}
{1+\tilde{\mathcal{K}}(\Delta)}=\frac{\tilde{\mathcal{K}}(\Delta)}{1+\tilde{\mathcal{K}}(\Delta)}
\frac{\Delta}{\tau_c(T)^2}\,\partial_T \tau_c(T).
\end{equation*}
This quantity describes how the normalized memory kernel changes with temperature at fixed delay.

The absorber is modeled as a single bosonic mode of frequency $\omega_a / 2\pi = 1$~GHz with mean thermal occupation $\bar n_a$.
For identical teeth, we define the dimensionless single-tooth dephasing strength \(g=(\lambda \tau)^2\).
We can rewrite, respectively, the one- and two-tooth coherence envelopes in Eqs.~\eqref{eq:SI_C1_WD} and~\eqref{eq:SI_C2_WD} as
\begin{equation}\label{eq:SI_C1_C2}
    C_j = e^{-\Gamma_{\phi_j}},\qquad j \in \{1, 2\},
\end{equation}
with \(\Gamma_{\phi_1}(T) = g\,\bar{n}_a(T)\) and 
\(\Gamma_{\phi_2}(\Delta, T) = 2g\,\bar{n}_a(T)[1 + \tilde{\mathcal{K}}(\Delta)]\).
We choose the dimensionless dephasing strength $g=0.05$ to maintain the probe in the weak dephasing regime. 

The corresponding QFI can be generally expressed as
\begin{equation}\label{eq:SI_F1_F2}
\mathcal{F}_j = \frac{(\partial_T C_j)^2}{1 - C_j^2}=\frac{(\partial_T \Gamma_{\phi_j})^2}{e^{2\Gamma_{\phi_j}} - 1}, \qquad j \in \{1, 2\},
\end{equation}
and the memory efficiency with two identical teeth is
\begin{equation}\label{eq:SI_A}
    \mathcal{A} = \frac{\mathcal{F}_2}{2\mathcal{F}_1}.
\end{equation}

Fig.~\ref{fig:process_kernel}(a) shows how the Lorentzian correlation kernel in Eq.~\eqref{eq:SI_Lor_K} evolves with $T$ and $\Delta$, highlighting the exponential decay of thermal photon-number correlations at each temperature: usable thermal memory decays more rapidly as the absorber is warmed.  
The white dashed curve marks the contour where 
\(\tilde{\mathcal{K}}=e^{-1}\), 
indicating the correlation-time boundary at which the kernel amplitude has dropped to $1/e$ of its zero-delay value.

Fig.~\ref{fig:process_kernel}(b) displays the pure amplitude gain factor $(1+\tilde{\mathcal{K}})$ appearing in Eq.~\eqref{eq:A_responsivity_factorized}, which quantifies the correlation-induced boost to the probe's phase variance.  
This correlated gain is broad and substantial at low temperature but narrows rapidly as the absorber's correlation time decreases with increasing $T$.

Fig.~\ref{fig:process_kernel}(c) shows the relative kernel responsivity 
$S_{\tilde{\mathcal{K}}} = \partial_T \tilde{\mathcal{K}}/(1+\tilde{\mathcal{K}})$,  
which determines how memory modifies the temperature derivative of the QFI.  
Following the $1/e$ dashed contour, where $|\partial_T \tilde{\mathcal{K}}|$ is maximized, $S_{\tilde{\mathcal{K}}}$ forms a pronounced negative ridge, revealing a strong competition with the population responsivity that can suppress the memory efficiency.

Fig.~\ref{fig:process_kernel}(d) presents the normalized two-tooth QFI \(\widetilde{\mathcal{F}}_2(T,\Delta)\), which inherits the non-monotonic structure of the kernel’s temperature responsivity: correlations enhance the QFI at short delays but eventually diminish as fluctuations decorrelate.

Fig.~\ref{fig:process_kernel}(e) shows the corresponding memory advantage $\mathcal{A}(\Delta)=\mathcal{F}_2/[2\mathcal{F}_1]$.  
Regions with $\mathcal{A}>1$ reflect correlation-assisted enhancement, while regions with $\mathcal{A}<1$ arise where the negative responsivity ridge in panel~(c) overwhelms the positive population slope.  
Green markers indicate the delay at which $\mathcal{A}$ reaches its minimum for selected temperatures.

Fig.~\ref{fig:process_kernel}(f) gives line cuts of $\mathcal{A}(\Delta)$ at three temperatures (15, 30, and 45 mK).  
At low temperature, the minimum of $\mathcal{A}$ is shifted away from the $1/e$ contour because $\tau_c(T)$ is nearly temperature-independent, causing $\partial_T \tilde{\mathcal{K}}$ to peak at a different delay.  
At higher temperatures, where $\tau_c(T)$ varies more strongly, the minimum aligns with the correlation-time boundary.

\paragraph*{Physical insight.} 
The non-monotonic behavior of the two-tooth QFI observed in Fig.~\ref{fig:process_kernel} can be understood as arising from a competition between two physically distinct sources of probe dephasing: the absorber’s mean thermal occupation and its temporal correlations.

We factorize the memory kernel Eq.~\eqref{eq:SI_Lor_K} as
\begin{equation*}
    \mathcal{K} = \bar{n}_a\, e^{-\Delta/\tau_c} = \bar{n}_a\, \tilde{\mathcal{K}},
\end{equation*}
where $\bar n_a$ is the absorber’s mean thermal occupation and $\tilde{\mathcal{K}}$ captures the normalized decay of population correlations through the correlation time $\tau_c(T)$.

Using the first-order approximation $e^{2\Gamma_{\phi_j}}-1 \simeq 2\Gamma_{\phi_j}$ in the weak-dephasing regime, Eq.~\eqref{eq:SI_F1_F2} can be rewritten as
\begin{equation*}
\mathcal{F}_j \simeq \frac{(\partial_T \Gamma_{\phi_j})^2}{2\,\Gamma_{\phi_j}},
\end{equation*}
in which 
\(\partial_T \Gamma_{\phi_1} = g\,\partial_T \bar{n}_a\) and 
\(\partial_T \Gamma_{\phi_2} = 2 g[(1+\tilde{\mathcal{K}})\,\partial_T \bar{n}_a + \bar{n}_a\,\partial_T \tilde{\mathcal{K}}]\). 

Thus, the memory efficiency Eq.~\eqref{eq:SI_A} is expressed as
\begin{equation*}
\begin{aligned}
\mathcal{A} &\simeq \frac{\Bigl[(1+\tilde{\mathcal{K}})\,\partial_T \bar{n}_a+\bar{n}_a\,\partial_T \tilde{\mathcal{K}}\Bigr]^2}{(1+\tilde{\mathcal{K}})(\partial_T \bar{n}_a)^2}\\
&= (1+\tilde{\mathcal{K}})\bigg[1+\frac{\partial_T\tilde{\mathcal{K}}/(1+\tilde{\mathcal{K}})}{\partial_T \bar{n}_a/\bar{n}_a}\bigg]^2 .
\end{aligned}
\end{equation*}
We introduce two dimensionless temperature responsivities:
\begin{equation*}
    S_{\bar{n}_a} = \partial_T \bar{n}_a / \bar{n}_a, \qquad
    S_{\tilde{\mathcal{K}}} = \partial_T \tilde{\mathcal{K}} / (1+\tilde{\mathcal{K}}),
\end{equation*}
and obtain the compact factorized form
\begin{equation} \label{eq:A_responsivity_factorized}
    \mathcal{A} \simeq (1+\tilde{\mathcal{K}})\,(1 + \frac{S_{\tilde{\mathcal{K}}}}{S_{\bar{n}_a}})^2.
\end{equation}

Equation~\eqref{eq:A_responsivity_factorized} makes the physics transparent: The prefactor $(1+\tilde{\mathcal{K}})\geq 1$ is a pure amplitude gain from correlations. It quantifies how much the correlated absorber fluctuations increase the total phase variance seen by the probe relative to the independent-tooth baseline.
The squared factor $(1 + S_{\tilde{\mathcal{K}}}/S_{\bar{n}_a})^2$ encodes a responsivity competition between the population slope $S_{\bar{n}_a}$ and the kernel slope $S_{\tilde{\mathcal{K}}}$. When $S_{\tilde{\mathcal{K}}}$ has the same sign as $S_{\bar{n}_a}$, correlations sharpen the temperature dependence of the dephasing and the QFI is further enhanced. When $S_{\tilde{\mathcal{K}}}$ has the opposite sign, the kernel partially cancels the population responsivity and correlations can suppress the QFI below the memoryless reference ($\mathcal{A}<1$).

In the Lorentzian kernel model used in Fig.~\ref{fig:process_kernel}, short delays $\Delta\ll \tau_c(T)$ exhibit $\tilde{\mathcal{K}}\approx 1$ and only weak kernel responsivity, $S_{\tilde{\mathcal{K}}}/S_{\bar n_a}\approx 0$, so that the amplitude-gain factor in Eq.~\eqref{eq:A_responsivity_factorized} contributes to the enhancement $\mathcal{A}>1$. Around the correlation--time contour $\Delta\simeq \tau_c(T)$, the kernel becomes strongly temperature--sensitive, and the sign and magnitude of \(S_{\tilde{\mathcal{K}}}/S_{\bar n_a}\) drive the competition factor below unity.  
This produces the observed memory disadvantage $\mathcal{A}<1$ even though the overall dephasing amplitude is larger than in the memoryless case.

Although we illustrate this behavior using a Lorentzian memory kernel $\mathcal{K}(\Delta)$, the structure of Eq.~\eqref{eq:A_responsivity_factorized} is generic: for any correlated noise source whose dephasing can be decomposed into single-time population term and a two-time kernel, the same competition between population responsivity $\partial_T \bar n_a / \bar n_a$ and correlation responsivity $\partial_T \tilde{\mathcal{K}} / (1 + \tilde{\mathcal{K}})$ will govern whether correlations enhance or suppress the protocol-level thermometric performance.

\section{Noise Spectrum Reconstruction}
Fig.~\ref{fig:noise_spectrum_discrimination} illustrates how different temporal correlation models for the absorber noise imprint distinct delay dependence in the two-tooth coherence and how the same data can be used to reconstruct the underlying noise spectrum $S_{nn}(\omega)$.

We work in dimensionless units and focus purely on the shape of the decay; all pre-factors are absorbed into a single variance scale. 
We write the single-tooth and correlation contributions as
\begin{equation*}
    A_b=\lambda^2(\tau_1^2+\tau_2^2)\,\bar n_a, \qquad
    A_c=2\lambda^2\tau_1\tau_2\,\bar n_a,
\end{equation*}
and Eq.~\eqref{eq:SI_C2_WD} becomes
\begin{equation}\label{eq:SI_Snn}
    C_2(\Delta)=\exp\!\big[-A_b - A_c\,\tilde{\mathcal{K}}(\Delta)\big].
\end{equation}

In the numerics we intentionally set \(\lambda = 0.5,~\tau_1=\tau_2=3 \tau_0,~\mathrm{and}~\mathrm{Var}(n_a)=1\) to maximize the contrast. We sample the delay on a logarithmic grid spanning many decades $10^{-3} \leq \Delta / \tau_0 \leq 10^{4}$, so that both very short- and very long-delay behavior are resolved.
We consider three representative forms of the normalized kernel $\mathcal{K}(\Delta)$ at a fixed temperature:

\paragraph*{White-noise-like (quasi-$\delta$-correlated) kernel:} Very short correlation time modeled by a narrow Gaussian,
\begin{equation*}
    \tilde{\mathcal{K}}_{\mathrm{white}} \propto e^{-(\Delta / \sigma_w)^2}, \qquad \sigma_w = 10^{-3},
\end{equation*}
normalized so that \(\tilde{\mathcal{K}}_{\mathrm{white}}(0)=1\).

\paragraph*{Lorentzian-correlated kernel:} Exponential decay with correlation time $\tau_c$,
\begin{equation*}
    \tilde{\mathcal{K}}_{\mathrm{Lor}} = e^{-\Delta / \tau_c}, \qquad \tau_c = 10,
\end{equation*}
again normalized so that \(\tilde{\mathcal{K}}_{\mathrm{Lor}}(0)=1\).

\paragraph*{$1/f$-like, heavy-tailed kernel:} Slow, algebraic decay designed to produce a broad low-frequency tail,
\begin{equation*}
\tilde{\mathcal{K}}_{1/f}=\frac{1}{1+(\Delta/\tau_f)^{\gamma}}, \qquad \tau_f = 0.1, \qquad 0<\gamma<1,
\end{equation*}
with $\gamma=0.6$ in Fig.~\ref{fig:noise_spectrum_discrimination}. This form decays more slowly than an exponential and therefore mimics a $1/f$-like spectrum over an intermediate frequency range.

For each kernel we compute the corresponding two-tooth coherence Eq.~\eqref{eq:SI_Snn}, and normalize by the asymptotic long-delay value
\begin{equation*}
    C_{2,\infty}^{(j)}=\lim_{\Delta\to\infty}C_2^{(j)}(\Delta), \quad j\in\{\mathrm{white},\mathrm{Lor},1/f\}
\end{equation*}
which removes the delay-independent single-tooth factor.  
The plotted visibility is therefore
\begin{equation*}
    \tilde{C}_2^{(j)}(\Delta) = e^{-A_c\,\tilde{\mathcal{K}}_j(\Delta)},
\end{equation*}
so that all curves start below unity at $\Delta=0$ and recover toward unity as the correlation term switches off.

For reconstruction, we invert the coherence–kernel relation:
\begin{equation*}
    \tilde{\mathcal{K}}_j^{\mathrm{(rec)}}(\Delta)=\frac{-\ln C_2^{(j)}(\Delta) - A_b}{A_c},
\end{equation*}
which coincides with the true $\mathcal{K}_j(\Delta)$ up to small numerical errors.  
Equivalently, if the long-delay-normalized visibility \(\tilde C_2^{(j)}=C_2^{(j)}/C_{2,\infty}^{(j)}\) is used, then \(\tilde{\mathcal{K}}_j^{\mathrm{(rec)}}(\Delta)=-\ln \tilde C_2^{(j)}(\Delta)/A_c\).
We then compute an estimate of the noise spectrum via a discrete cosine transform,
\begin{equation*}
    S_{nn}^{\mathrm{(rec)}}(\omega) \propto \int_{0}^{\infty}\!d\Delta\,\tilde{\mathcal{K}}_j^{\mathrm{(rec)}}(\Delta)\,\cos(\omega\Delta),
\end{equation*}
which is implemented numerically as
\begin{equation*}
    S_{nn}^{\mathrm{(rec)}}(\omega_i)=2\int_{0}^{\Delta_{\max}}\!d\Delta\,\tilde{\mathcal{K}}_j^{\mathrm{(rec)}}(\Delta)\,\cos(\omega_i\Delta),
\end{equation*}
for a logarithmic grid of frequencies $10^{-4} \le \omega/2\pi \le 10^{2}$.  
The reconstructed spectra are rescaled so that their maxima are normalized to unity.

We compare three representative models of absorber noise and demonstrate how the corresponding two-tooth coherence encodes their spectral structure.

Panel~(a) shows normalized two-tooth coherence $\tilde{C}_2(\Delta)$ (visibility) versus delay~$\Delta$ on a logarithmic axis for three kernel models:  
quasi-white noise (narrow, rapidly decaying kernel), Lorentzian-correlated noise (exponential kernel), and a $1/f$-like heavy-tailed kernel.  
All traces are normalized by their long-delay values so that they share the same independent-tooth baseline.  
The white-noise curve recovers rapidly and saturates, reflecting an extremely short correlation time.  
The Lorentzian curve recovers more gradually, while the $1/f$-like curve displays the slowest, almost power-law-like recovery, signaling long-lived correlations extending over many decades in~$\Delta$.
 
Panel~(b) shows reconstructed spectra $S_{nn}^{\mathrm{(rec)}}(\omega)$ obtained by inverting the coherence to recover $\tilde{\mathcal{K}}(\Delta)$ and applying a cosine transform.  
Each curve is normalized to its own maximum.  
The reconstructed white-noise spectrum is nearly flat over a broad intermediate band, with a high-frequency roll-off set by the finite time resolution and a low-frequency roll-off set by the finite maximum delay.  
The Lorentzian kernel yields a reconstructed spectrum that is peaked at low frequency and decays approximately as $1/(1+\omega^2\tau_c^2)$.  
The $1/f$-like kernel produces a spectrum that approximates a $1/\omega$ dependence across several decades, again with cutoffs at the extremal frequencies imposed by the finite $\Delta$-window and sampling.  
Together, these curves demonstrate that the two-tooth visibility encodes enough information to distinguish qualitatively different temporal noise structures via simple Fourier inversion.

\section{Experimental implementation and robustness}\label{app:probe_coherence}
The TBQC protocol builds on the coherence-mediated circuit-QED thermometry architecture of Ref.~\cite{zhu2025coherence}, which consists of four hardware modules: a thermal absorber mode, a long-lived bosonic probe mode, an engineered cross-Kerr coupling element, and a qubit-assisted readout module for the final probe coherence.

A high-coherence 3D cavity can serve as the bosonic probe memory, while a lower-frequency resonator or terminated transmission-line mode acts as the thermal absorber~\cite{romanenko2020three,reagor2013reaching,kim2025ultracoherent,zhu2022high,crowley2023disentangling}.
The absorber--probe cross-Kerr interaction can be generated by an intermediate transmon or nonlinear coupler, including fixed-frequency or tunable-coupler implementations~\cite{gao2019entanglement,miano2022frequency,maiti2025linear,kounalakis2018tuneable,didier2015fast}. 
Finally, a high-coherence superconducting qubit dispersively coupled to the probe provides the Ramsey/readout module used to convert the final probe coherence into a measurable signal~\cite{krantz2019quantum,bal2024systematic}.
Relative to Ref.~\cite{zhu2025coherence}, the additional requirement is therefore not a new hardware element, but the preservation and readout of probe coherence over a controlled inter-tooth delay.

The two-tooth protocol does not require time-bin encoding, delayed feedback, or measurement conditioning. 
A ``tooth'' is a calibrated time window during which the effective cross-Kerr coupling is active.
A representative pulse sequence is:
\begin{enumerate}[nosep]
    \item Prepare the probe mode in a coherent state \(|\alpha\rangle\), and let the absorber mode thermalize at temperature \(T\).
    \item Activate the effective cross-Kerr interaction for a calibrated duration \(\tau_1\).
    \item Suppress the effective interaction and let the absorber evolve freely for a delay \(\Delta\), while the probe stores the phase imprinted during the first tooth.
    \item Activate the interaction again for a duration \(\tau_2\).
    \item Read out the final probe coherence using the dispersively coupled qubit.
\end{enumerate}

\subsection{Intrinsic probe dephasing}
While cavity experiments often report photon storage lifetimes rather than separately extracted pure-dephasing times, millisecond- to second-scale lifetimes have been demonstrated in 3D superconducting cavities~\cite{romanenko2020three,kim2025ultracoherent,reagor2013reaching}.
These values are well above the microsecond-scale tooth durations and delays considered in the present work, leaving a favorable coherence budget for the TBQC sequence.

We consider a long-lived probe with finite intrinsic dephasing time \(T_\varphi^p\), and model the corresponding probe envelope as
\begin{equation*}
C_p(t)=e^{-\Gamma_\varphi^p t},\qquad
\Gamma_\varphi^p=1/T_\varphi^p .
\end{equation*}
Probe dephasing only multiplies the measured probe coherence by an independently calibratable envelope.

The absorber number fluctuations are described by the stationary kernel
\begin{equation*}
\mathcal{K}(\Delta)=\big\langle \delta n_a(t)\,\delta n_a(t+\Delta)\big\rangle_T .
\end{equation*}
This kernel is set by the absorber dynamics and temperature, and is not modified by intrinsic probe dephasing.

The one-tooth coherence visibility is corrected as
\begin{equation*}
C_1^{(j),\mathrm{full}}(\tau_j)=C_p(\tau_j)\,C_1^{(j)}(\tau_j).
\end{equation*}
In the short-tooth limit \(\tau_j\ll T_\varphi^p\), this correction is typically negligible for each individual interaction window.

For the two-tooth sequence, the measured visibility can be written as
\begin{equation}
C_2^{\mathrm{full}}(\Delta)=C_p(T_s)\,C_2(\Delta),\quad
T_s=\tau_1+\Delta+\tau_2,
\label{eq:SI_modified_C2}
\end{equation}
where \(C_2(\Delta)\) is the absorber-induced coherence envelope derived in Eq.~\eqref{eq:SI_C2_WD}. 
The probe envelope \(C_p(T_s)\) is a multiplicative contrast factor that can be calibrated independently. 
Thus intrinsic probe dephasing attenuates the measured visibility but does not generate the absorber-specific delay dependence of \(\mathcal{K}(\Delta)\).

Substituting Eq.~\eqref{eq:SI_modified_C2} into the QFI expression and assuming \(C_p\) is temperature independent over the operating range gives
\begin{equation}
\mathcal{F}_2^\mathrm{full}(\Delta)=\frac{C_p^2(T_s)\,\big[\partial_T C_2(\Delta)\big]^2}{1-C_p^2(T_s)\,C_2^2(\Delta)}.
\label{eq:SI_modified_F}
\end{equation}
The absorber derivative entering the numerator, including the memory contribution through \(\partial_T\mathcal{K}(\Delta)\), is unchanged by intrinsic probe dephasing. The practical effect of \(C_p\) is to reduce the available contrast and therefore the measurable QFI.

\subsection{Experimental feasibility}
The numerical examples in the main text use a representative circuit-QED parameter regime rather than a device-specific optimization. 
We take an absorber frequency \(\omega_a/2\pi=1~{\rm GHz}\), corresponding to \(\hbar\omega_a/k_B\simeq48~{\rm mK}\), and temperatures \(T=10\)--\(50~{\rm mK}\). 
The plotted memory kernel is phenomenological: it is parameterized by an absorber population-noise correlation time \(\tau_c(T)\), which captures the crossover from a slow low-temperature regime to a more rapidly decorrelating high-temperature regime. 
In the simplest Markovian realization, this memory time is set by the absorber relaxation rate, \(\tau_c\simeq \kappa_a^{-1}\), while in a device with additional thermal bottlenecks or slow fluctuators it should be interpreted as an effective correlation time to be measured independently.

In an experiment, the relevant parameters are the absorber--probe cross-Kerr rate \(\lambda\), the tooth durations \(\tau_j\), the inter-tooth delay \(\Delta\), the absorber relaxation rate \(\kappa_a\), the probe coherence times \(T_1^p\) and \(T_\varphi^p\), and the qubit--probe dispersive pull \(\chi\) used for the final readout. 
For a single-pole absorber relaxation model, \(\kappa_a\) determines the population-correlation time, \(\tau_c\simeq\kappa_a^{-1}\). 
The short-tooth approximation requires \(\tau_j\ll\tau_c\), so that each tooth samples an approximately fixed absorber occupation. 
The delay \(\Delta\) is then scanned over the scale of this correlation time: for \(\Delta\ll\tau_c\), the two teeth sample nearly the same absorber fluctuation, while for \(\Delta\gg\tau_c\), the sequence approaches the independent-tooth benchmark.

The cross-Kerr phase per tooth should be large enough to produce a resolvable visibility change, while remaining small enough to avoid complete contrast loss. 
In the weak-dephasing limit, the correlated part of the two-tooth visibility scales as
\begin{equation*}
\Delta C_2(\Delta)\sim2\lambda^2\tau_1\tau_2|\mathcal{K}(\Delta)|,
\end{equation*}
up to known protocol-dependent prefactors. 
A useful experimental resolvability condition is therefore
\begin{equation*}
2\lambda^2\tau_1\tau_2|\mathcal{K}(\Delta)|\gtrsim\delta C_{\rm meas},
\end{equation*}
where \(\delta C_{\rm meas}\) is the uncertainty in the measured visibility. 
For projection-noise-limited readout with \(N_{\rm sh}\) repetitions and readout contrast \(\mathcal{V}_{\rm ro}\),
\begin{equation*}
\delta C_{\rm meas}\sim\frac{1}{\mathcal{V}_{\rm ro}\sqrt{N_{\rm sh}}}.
\end{equation*}
Equivalently, for a shot rate \(r_{\rm sh}\) and averaging time \(t_{\rm av}\), one has \(N_{\rm sh}=r_{\rm sh}t_{\rm av}\).

The qubit--probe dispersive pull \(\chi\) sets the speed and contrast of the final coherence readout. 
It must allow a calibrated Ramsey-type mapping of the final probe coherence within a readout window short compared with the relevant probe memory time. 
This preserves the separation of roles emphasized in Ref.~\cite{zhu2025coherence}: the bosonic probe stores the phase information throughout the two-tooth sequence, whereas the qubit is used only as a final readout resource.

We summarize the representative operating conditions for a circuit-QED implementation of the two-tooth protocol in Table~\ref{tab:exp_window}.

\begin{table}[t]
\caption{Representative operating window for a circuit-QED implementation of the two-tooth protocol. The first two rows specify the illustrative thermal regime; the remaining rows summarize the hierarchy needed to resolve the absorber two-time population correlator.}
\label{tab:exp_window}
\begin{ruledtabular}
\begin{tabular}{ll}
Quantity / role & Representative condition \\
\hline
Absorber mode & \(\omega_a/2\pi \sim 1~{\rm GHz}\) \\
Thermal regime & \(T\sim 10{-}50~{\rm mK}\) \\
Corr. time & \(\tau_c\simeq \kappa_a^{-1}\) \\
Tooth duration & \(\tau_j\ll \tau_c\) \\
Inter-tooth delay & \(0\lesssim \Delta \lesssim {\rm few}\,\tau_c\) \\
Kerr phase & \(\lambda\tau_j\bar n_a\lesssim 1\) \\
Probe coherence & \(T_\varphi^p,T_1^p \gtrsim \tau_1+\Delta+\tau_2\) \\
Qubit readout & calibrated dispersive mapping \\
Shot noise & \(\delta C_{\rm meas}\sim(\mathcal{V}_{\rm ro}\sqrt{N_{\rm sh}})^{-1}\) \\
\end{tabular}
\end{ruledtabular}
\end{table}

\subsection{Robustness and calibration}

The central experimental challenge is to separate the absorber memory signal from contrast loss caused by the probe, coupler, and readout over the full two-tooth sequence. 
This separation can be achieved with a set of reference measurements. 
First, an absorber-decoupled or far-detuned sequence measures a reference contrast envelope \(C_{\rm ref}(T_s)\), with \(T_s\simeq \tau_1+\Delta+\tau_2\). 
This envelope captures intrinsic probe decay, residual coupler-induced dephasing, and finite readout contrast, but excludes absorber population correlations. 
Second, single-tooth measurements calibrate the normalized one-tooth visibility factors \(C_1^{(j)}\) and the effective Kerr pulse areas. 
Third, measurements at large delay, \(\Delta\gg \tau_c\), establish the independent-tooth baseline, where \(\mathcal{K}(\Delta)\to0\).

Using a high-frequency probe, e.g. \(\omega_p/2\pi\sim9~{\rm GHz}\), strongly suppresses spurious thermal population in the readout channel. 
The dominant probe-channel imperfections are therefore better treated as contrast loss, excess dephasing, and finite readout efficiency, rather than as thermal occupation of the probe mode. 
In the qubit-assisted Ramsey readout considered here, these effects renormalize the measured visibility and increase the required averaging time, but they do not by themselves generate the absorber-specific delay dependence expected from \(\mathcal{K}(\Delta,T)\).

During each tooth window, the tunable coupling element can introduce additional probe dephasing through flux noise, pump fluctuations, or residual nonlinear loss~\cite{miano2022frequency, maiti2025linear}. 
This contribution should be calibrated under control conditions as close as possible to the sensing sequence. 
A clean calibration can be obtained using a reference probe--coupler module driven with the same coupler waveform, tooth durations, inter-tooth delay, and readout sequence. 
The resulting visibility defines a reference envelope \(C_{\rm ref}(T_s)\), with \(T_s\simeq\tau_1+\Delta+\tau_2\), which captures probe decay, coupler-induced dephasing, and readout contrast.

In addition, the coupler may also add noise or heating directly to the absorber, which changes the effective absorber fluctuations being measured and should be treated as a systematic modification of the measured absorber kernel. 
This backaction can be bounded by repeating the extraction while varying the coupler bias, pump power, and tooth duration, and by comparing the resulting kernels with single-tooth measurements and the large-delay independent-tooth baseline. 
A genuine absorber-memory signal should remain consistent after calibration of the effective Kerr pulse area, whereas coupler-induced absorber heating or backaction is expected to show a stronger dependence on the coupler operating point.

Timing errors of the Kerr teeth enter in two distinct ways. 
Fluctuations in the tooth duration produce pulse-area jitter and therefore renormalize the effective Kerr areas. 
This effect can be absorbed into a calibrated area product \(A_{12}^{\rm eff}\), obtained from single-tooth measurements as a function of the programmed pulse duration. 
Jitter in the relative placement of the two teeth changes the effective delay, \(\Delta\to\Delta+\epsilon\), and broadens the measured kernel over the timing-response distribution. 
When the delay jitter is small compared with the absorber correlation time, \(\sigma_\Delta\ll\tau_c\), this broadening is perturbative; otherwise, the independently measured timing response should be included in the kernel fit.

With these calibrations, the corrected two-tooth visibility is \(C_2^{\rm corr}(\Delta) = C_2^{\rm full}(\Delta) / C_{\rm ref}(T_s).\)
In the weak-dephasing regime, the absorber kernel is extracted from the residual delay-dependent visibility as
\[\mathcal{K}_{\rm meas}(\Delta) = -\frac{1}{2A_{12}^{\rm eff}}\ln\!\left[\frac{C_2^{\rm corr}(\Delta)}{C_1^{(1)}(\tau_1)C_1^{(2)}(\tau_2)}\right],\]
where \(C_1^{(j)}\) are the calibrated single-tooth visibilities. 
The large-delay limit provides a direct consistency check,
\[C_2^{\rm corr}(\Delta\gg\tau_c)\rightarrow C_1^{(1)}(\tau_1)C_1^{(2)}(\tau_2).\]
Slow drifts in temperature, pump amplitude, or readout gain can be further suppressed by interleaving different delay values, including the large-delay reference. 
Thus the main imperfections either reduce visibility, renormalize calibrated pulse areas, broaden the delay resolution, or modify the effective absorber kernel; the absorber-memory contribution is isolated by the combined use of reference-envelope, single-tooth, and large-delay calibrations.

\section{General \(m\)-tooth probe sequence}\label{app:m_tooth}
\begin{figure*}[t!]
    \centering
    \includegraphics[width=0.99\linewidth]{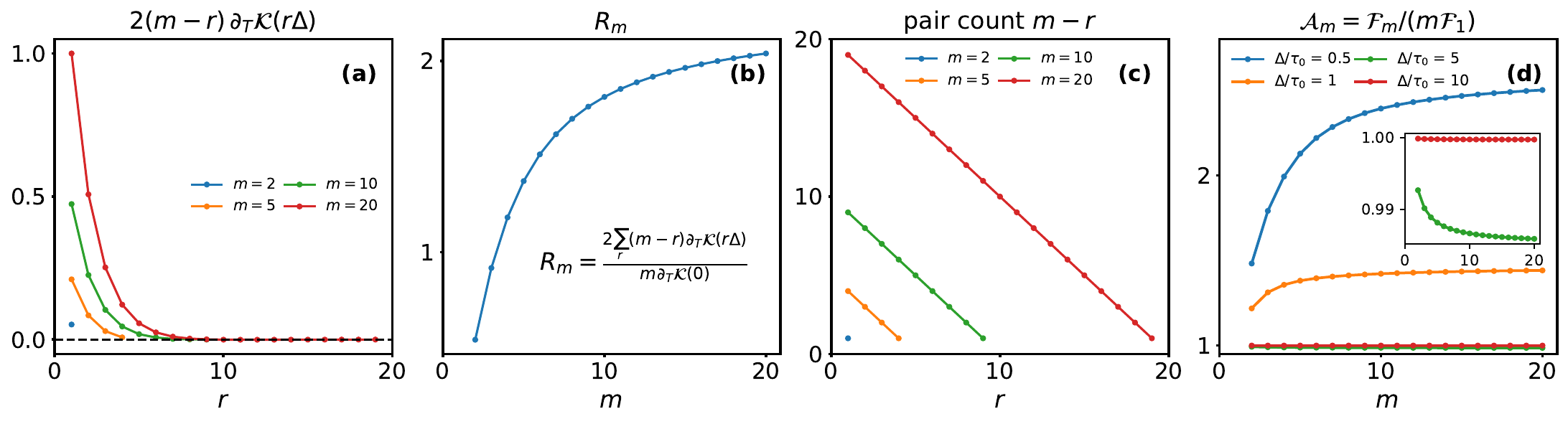}
    \caption{Reshaping of correlation responsivity in an \(m\)-tooth comb.
    (a) Weighted memory responsivity \(2(m-r)\partial_T\mathcal{K}(r\Delta)\) as a function of delay index \(r\), showing how larger combs collect contributions from multiple pair separations.
    (b) Integrated memory responsivity \(R_m=2\sum_{r=1}^{m-1}(m-r)\partial_T\mathcal{K}(r\Delta)/[m\partial_T\bar n_a]\), normalized by the instantaneous response.
    (c) Pair multiplicity \(m-r\), i.e., the number of tooth pairs separated by \(r\Delta\).
    (d) Memory efficiency \(\mathcal{A}_m=\mathcal{F}_m/(m\mathcal{F}_1)\) for representative delays. For very large delays the independent-tooth value \(\mathcal{A}_m\simeq1\) is recovered, while finite delays can yield either enhancement or a small correlation-induced penalty depending on the sign of the memory responsivity.
    The inset magnifies the near-unity region to show the approach to the independent-tooth limit.}
    \label{fig:m_tooth_reshaping}
\end{figure*}

In the main text we focused on one- and two-tooth probe sequences. 
Here we show that the same formalism extends to a general \(m\)-tooth sequence. 
This clarifies that the two-tooth protocol is the minimal, nontrivial member of a broader temporal-comb hierarchy: one tooth probes only equal-time fluctuations, while two separated teeth are sufficient to expose a connected two-time absorber correlation.

We consider \(m\) interaction windows, labeled by \(j=1,\ldots,m\), with durations \(\tau_j\) and center times \(t_j\). 
During these windows the effective cross-Kerr interaction is
\begin{equation*}
H_{\rm int}/\hbar=\lambda \hat n_a\hat n_b .    
\end{equation*}
For the coherence element probed in the main text, the absorber-induced factor is
\begin{equation}
C_m=\left\langle\exp\left[-i\lambda\int dt\, Y_m(t)\hat n_a(t)\right]\right\rangle ,
\end{equation}
where \begin{equation*}Y_m(t)=\sum_{j=1}^m f_j(t-t_j)\end{equation*} is the switching function of the comb. 
The function \(f_j\) is localized around the \(j\)th tooth and satisfies \(\int dt\,f_j(t)=\tau_j\).

Writing \(\hat n_a(t)=\bar n_a+\delta \hat n_a(t)\), the deterministic mean phase can be removed by demodulation. 
The remaining coherence envelope is governed by the connected cumulants of the absorber population. 
In the weak-dephasing regime, the coherent-state visibility is \(C_m\simeq e^{-\Gamma_{\phi_m}}\), and the effective coherence dephasing rate is written as
\begin{equation}\label{eq:m-tooth dephase}
   \Gamma_{\phi_m}=\lambda^2\int dt\,dt'\,Y_m(t)Y_m(t')\,\mathcal{K}(t-t'),
\end{equation}
where \(\mathcal{K}(t-t')=\langle \delta \hat n_a(t)\delta \hat n_a(t')\rangle_T.\)

For short teeth, where \(\mathcal{K}\) varies slowly over each interaction window, Eq.~\eqref{eq:m-tooth dephase} becomes
\begin{equation}
\Gamma_{\phi_m}\simeq\lambda^2\sum_{i,j=1}^{m}\tau_i\tau_j\mathcal{K}(t_i-t_j),
\end{equation}
In the \(\bar n_a \ll 1\) limit, the diagonal single-tooth contribution $\mathcal{K}(0) = \mathrm{Var}(n_a)\approx \bar n_a$. 
Equivalently,
\begin{equation}
\Gamma_{\phi_m}\simeq\lambda^2\left[\sum_{j=1}^{m}\tau_j^2\bar n_a+2\sum_{i<j}\tau_i\tau_j\mathcal{K}(t_i-t_j)\right].    
\end{equation}
The first term is the sum of independent single-tooth dephasing contributions, while the second term contains all pairwise temporal correlations between distinct teeth. 
Thus, increasing \(m\) increases the number of correlation-sensitive pair terms from one in the two-tooth protocol to \(m(m-1)/2\).

The one-tooth result follows by setting \(m=1\),
\begin{equation*}
\Gamma_{\phi_1}\simeq\lambda^2\tau_1^2\bar n_a,    
\end{equation*}
whereas the two-tooth result is
\begin{equation*}
\Gamma_{\phi_2}\simeq\lambda^2\Big[(\tau_1^2+\tau_2^2)\bar n_a+2\tau_1\tau_2\mathcal{K}(\Delta)\Big].
\end{equation*}
This is the minimal sequence in which a genuine two-time correlator enters the probe coherence.

The temperature response of the general comb is
\begin{equation*}
\partial_T\Gamma_{\phi_m}\simeq\lambda^2\left[\sum_{j=1}^{m}\tau_j^2\partial_T\bar n_a+2\sum_{i<j}\tau_i\tau_j\partial_T\mathcal{K}(t_i-t_j)\right].    
\end{equation*}
This expression makes clear how the competition discussed in the main text generalizes: the instantaneous responsivity is carried by \(\partial_T\bar n_a\), while the delay-dependent correlation responsivity is replaced by a weighted sum over all pair delays.

For equally spaced identical teeth, \(t_j=t_0+j\Delta\) and \(\tau_j=\tau\),
\begin{equation}\label{eq:Gamma_M_equal}
\begin{aligned}
\partial_T\Gamma_{\phi_m}\simeq\lambda^2\tau^2\Big[m\,\partial_T\bar n_a +2\sum_{r=1}^{m-1}(m-r)\partial_T\mathcal{K}(r\Delta)\Big].
\end{aligned}
\end{equation}
The \(m\)-tooth comb therefore does not simply repeat the two-tooth measurement. 
It reshapes the balance between equal-time response and temporal memory through the pair-weighting factors \(m-r\). 
Slowly decaying memory can be amplified by the many correlated tooth pairs, while rapidly decaying or sign-changing responsivity can be averaged out or partially canceled.

Equations~\eqref{eq:Gamma_M_equal} show explicitly how higher-tooth sequences reshape the memory contribution. 
In the two-tooth case, there is only one off-diagonal pair and the correlation responsivity is proportional to \(\partial_T\mathcal{K}(\Delta)\). 
For \(m>2\), this single term is replaced by the weighted sum \(2\sum_{r=1}^{m-1}(m-r)\partial_T\mathcal{K}(r\,\Delta)\). 
The factor \(m-r\) counts how many tooth pairs are separated by \(r\,\Delta\): nearest-neighbor pairs are most numerous, while longer-delay pairs are progressively fewer and are further suppressed by the decay of \(\mathcal{K}(r\,\Delta)\). 
Thus increasing \(m\) can amplify slowly decaying memory, but it also broadens the temporal averaging over many pair delays.

The protocol-level envelope QFI is expressed as 
\begin{equation}
\mathcal{F}_m=\frac{e^{-2\Gamma_{\phi_m}}(\partial_T\Gamma_{\phi_m})^2}{1-e^{-2\Gamma_{\phi_m}}}
\end{equation}
This is the QFI of the reduced output probe state for the specified comb sequence and demodulated coherence-envelope response. 
It is not a global optimization over arbitrary comb protocols or joint probe--absorber measurements. 
The two-tooth protocol is therefore not claimed to be globally optimal; it is the minimal correlation-resolved sequence, while higher-tooth combs provide systematic extensions for spectral selectivity and temporal-correlation reconstruction.

The potential advantage of increasing \(m\) is therefore not a universal improvement in sensitivity, but a controlled reshaping of the temporal filter applied to the absorber correlations. 
Larger combs contain \(m(m-1)/2\) off-diagonal tooth pairs and can therefore amplify slowly decaying memory contributions or provide greater selectivity to the correlation kernel through the choice of \(m\), \(\Delta\), and the tooth weights. 
This can improve the thermometric Fisher information when the weighted correlation responsivity has the same sign as the instantaneous responsivity. 
The corresponding experimental challenges are also clear from the same formulas: the total sequence time \(T_s\simeq m\tau+(m-1)\Delta\) increases with \(m\), making the protocol more sensitive to probe loss and dephasing; the accumulated Kerr phase can drive the visibility into a low-contrast regime; and timing jitter, pulse-area errors, and coupler-induced dephasing must be calibrated across more teeth. 
In addition, because the memory contribution is a weighted sum over many delays, rapidly decaying or sign-changing kernels can lead to partial cancellation and diminishing returns. 
Thus the \(m\)-tooth comb should be viewed as a tunable correlation-spectroscopy extension of the minimal two-tooth protocol, with an optimal \(m\) set by the absorber memory time, available probe coherence, and calibration accuracy.

Figure~\ref{fig:m_tooth_reshaping} visualizes this pair-weighting structure for equally spaced teeth. 
Panel~(a) shows the weighted memory responsivity \(2(m-r)\partial_T\mathcal{K}(r\Delta)\), illustrating how larger combs collect contributions from multiple pair delays while remaining dominated by the shortest correlated separations. 
Panel~(b) plots the integrated memory responsivity normalized by the instantaneous contribution,
\begin{equation*}
R_m = \frac{2\sum_{r=1}^{m-1}(m-r)\partial_T\mathcal{K}(r\Delta)}{m\,\partial_T\bar n_a},
\end{equation*}
showing that the correlation response saturates rather than increasing indefinitely with \(m\). 
Panel~(c) displays the pair-count factor \(m-r\), making clear that the multi-tooth comb reshapes the competition through pair combinatorics, not by simply repeating independent one-tooth measurements.
Panel~(d) plots the multi-tooth memory efficiency \(\mathcal{A}_m=\mathcal{F}_m/(m\mathcal{F}_1)\) for several inter-tooth delays. 
As the delay is increased, the comb response crosses over from constructive memory-assisted accumulation at short delay, to a finite-delay regime where the correlation responsivity can suppress the Fisher information, and finally to the Markovian independent-tooth limit \(\mathcal{A}_m\to1\) at large delay.

\bibliography{TBQC_Main}

\end{document}